# Transmission-eigenchannel velocity and diffusion

Azriel Z. Genack[1,2], Yiming Huang[1,2,3], Asher Maor[1,2,4] and Zhou Shi[1,2,5]

[1]Department of Physics, Queens College of the City University of New York, Flushing, New York 11367, USA
[2]Physics Program, The Graduate Center of the City University of New York, New York New York, 10016, USA
[3]Jinhua No.1 High School, Zhejiang, 321000, China
[4]Kent Optronics, Inc., Hopewell Junction, New York 12533, USA
[5]OFS Labs, 19 School House Road, Somerset, New Jersey 08873, USA

**Abstract**
The diffusion model is used to calculate the time-averaged flow of particles in stochastic media and the propagation of waves averaged over ensembles of disordered static configurations. For classical waves exciting static disordered samples, such as a layer of paint or a tissue sample, the flux transmitted through the sample may be dramatically enhanced or suppressed relative to predictions of diffusion theory when the sample is excited by a waveform corresponding to a transmission eigenchannel. Even so, it is widely acknowledged that the velocity of waves is irretrievably randomized in scattering media. Here we demonstrate in microwave measurements and numerical simulations that the statistics of velocity of different transmission eigenchannels remain distinct on all length scales and are identical on the incident and output surfaces. The interplay between eigenchannel velocities and transmission eigenvalues determines the energy density within the medium, the diffusion coefficient, and the dynamics of propagation. The diffusion coefficient and all scattering parameters, including the scattering mean free path, oscillate with the width of the sample as the number and shape of the propagating channels in the medium change.

**Introduction**
The diffusion equation describes the flow of particles, waves and energy from neutrons, electrical charge, molecules and microscopic particles to light, sound, and heat[1–4]. The diffusion model begins with the assumption that scattering is local—the velocity, **v**, is randomized within a distance of the transport mean free path, $\ell$, and is determined solely by scattering within the medium and not by its overall dimensions[1–4]. The average over time of the flux within the medium is determined by Fick's first law, $\boldsymbol{j} = -D\boldsymbol{\nabla}u$, where $\boldsymbol{j}$ is the current density, $D$ is the diffusion coefficient, and $u$ is the average particle concentration or energy density[2,3]. For particles in $d$ dimensions, the Boltzmann diffusion coefficient for particles is $D_B = \frac{1}{d}\text{v}\ell$ [1–4]. The diffusing quantity drops towards open boundaries and extrapolates to zero at a distance $z_b$ beyond the sample, which is proportional to $\ell$[2,5–9].

The diffusion model can also be applied to the average of propagation over random configurations in mesoscopic media, in which multiply scattered waves are temporally coherent throughout the medium[10–16]. The interference of classical and quantum mechanical waves in mesoscopic samples produces a stable speckle pattern of energy or particle density. The spatial field distribution has a correlation length of half the wavelength, $\lambda/2$ [17,18], and provides a fingerprint of the wave interaction with the material. Averaging such speckle patterns over an ensemble of random sample configurations yields a smooth profile of particle or energy density[16]. In the limit in which the probability that randomly scattered wave trajectories with width $\lambda/2$, known as Feynman paths, loop back upon a typical coherence length along the trajectory tends to zero[11,12,16], the average profile is a solution of the diffusion equation[2] with boundary conditions given in terms of $z_b$[2,6-9]. Photon diffusion in static samples leads to the inverse scaling of total optical transmission with sample thickness[19] in accord with Ohm's law, and to diffusive pulse propagation[20–22].

However, as the scattering strength of the medium or the confinement of the wave increases, the transmission and the diffusion coefficient are increasingly suppressed by the interference of waves crossing back upon themselves within the medium. When the probability that a Feynman path will cross a typical coherence length along the path approaches unity the wave becomes localized[11]. Transport is then suppressed relative to predictions of diffusion theory and quasi-normal modes of the medium become exponentially localized instead of being extended over the entire sample[23–25]. Thus, propagation in multiply scattering samples is diffusive for samples with lengths, $L$, for which $\ell < L < \xi$, where $\xi$ is the localization length.

The limits of diffusive propagation may also be given in terms of the ensemble average of the dimensional conductance, $g$, which is equivalent to the average of the classical transmittance, $g = \langle T \rangle$[26,27]. The dimensionless conductance is the average electronic conductance in units of the quantum of conductance, $\frac{e^2}{h}$, while the transmittance is the sum over all pairs of flux transmission coefficients between the $N$ incident and outgoing channels of the sample, $T = \sum_{a,b}^{N} |t_{ba}|^2 = \langle \text{Tr}(tt^\dagger) \rangle$[10,28,29]. Here, $t$ is the transmission matrix (TM) with elements $t_{ba}$. The TM is most often applied to the quasi-1D wire or waveguide geometry with constant cross section and reflecting sides. A natural choice for the channels is the set of the $N$ propagating modes of the empty waveguide. The eigenvalues of $tt^\dagger$ are the transmission eigenvalues, $\tau_n$, so that $T = \sum_{n=1}^{N} \tau_n$, with $\tau_n$ decreasing for increasing $n$. For $g > 1$, waves in multiply scattering media are diffusive with $g$ open transmission eigenchannels (TEs) with $\tau_n > \frac{1}{e}$[10,30], while for $g < 1$, transmission is small in all channels and waves are localized. Dorokhov showed that each of the $N$ TEs of a conducting wire scales differently with its own localization length[10]. Since $g$ falls below its maximum value of $N$ in the presence of scattering, waves propagate diffusively for $N \geqslant g \geqslant 1$.

The TEs are the singular vectors found in the singular value decomposition (SVD) of the TM, $t = \mathcal{U}\Lambda\mathcal{V}^\dagger$[10,27,30–34]. Here $\mathcal{V}$ and $\mathcal{U}$ are unitary matrices whose columns are the singular vectors on the input and output of the sample, respectively, and $\Lambda$ is a diagonal matrix whose elements are the singular values, $\lambda_n = \sqrt{\tau_n}$. The amplitudes of the $m^{th}$ channel in the $n^{th}$ TE on the input and output surfaces are $v_{nm}$ and $u_{nm}$, respectively. Here, the channels will be taken to be the waveguide modes.

Aside from the suppression of transport due to Anderson localization[23,35] and weak-localization precursors[11–16], dramatic deviations from diffusion theory arise in mesoscopic samples with $g > 1$ due to global correlation which produces strong variation of transmission in different TEs[10,30,36,32,37,33]. Transmission may be perfect[10,30,38–41] or vanish[42,43], and the energy within the sample may be greatly enhanced or suppressed relative to the diffusive solution[44–46] when the sample is excited by a TE. This makes it possible to control the transmission of classical waves [38,47–49,39,34,50,40,42].

The enhancement of transmission in highly transmitting TEs may be exploited, for example, to reduce the power required for cellular communication[40], while the enhancement of energy within a medium holds promise for medical imaging and intervention[34,50]. On the other hand, the suppression of transmission may enable high dynamic range switching and extreme sensitivity to sample deformation[51,47,49,34,42,50]. In random slabs thinner than the transport mean free path, $L < \ell$, the grain sizes of optical speckle patterns of different TEs differ[52]. This provides an approach towards engineering speckle correlation in thin samples for improved resolution for structured illumination microscopy[53].

In this study, we show that the velocity distributions of TEs on the input and output surfaces of multiply scattering media are not randomized even by multiple scattering. We focus on the longitudinal components of the transmission eigenchannel velocities (EVs), $v_n$, which are the weighted averages over the angular distribution of the velocity component of the wave normal to the sample surface for different TEs. In the waveguide geometry this may be computed as the weighted averages over the distributions of group velocities of waveguide modes. The statistics of the $v_n$ and their ensemble averages are the same on both sides of the waveguide on all length scales and they approach different values as the sample length increases on length scales of a few transport mean free paths. As a result, the different TEs have different speckle patterns on the sample surface.

The interplay between the $\tau_n$ and $v_n$ yields the energy density on the open surfaces of the sample, as well as $D$ and $z_b$. As a result, $g$ and $D$ scale differently. These parameters, as well as the scattering mean free path, $\ell_s$, in which spatial coherence is lost, vary with the width of the sample as the number and shape of the transverse propagating modes change. The impact of transverse boundaries on the Boltzmann diffusion coefficient and the Thouless conductance is given.

**Results**
**Measurement of eigenchannel velocities.** Microwave measurements of spectra of the in- and out-of-phase components of field transmission coefficients, are described in Methods and in Supplementary Fig. 1(a). Spectra are obtained for two perpendicular orientations of wire antennas on the input and output of the sample on a square grid of points with use of a vector network analyser, as shown in Supplementary Fig. 1(b)[48]. The sample is composed of randomly positioned dielectric elements contained in a copper tube. Over the frequency range of the experiment of 14.70–14.94 GHz, the number of propagating modes supported by the waveguide changes from $N = 62$ to 63. The wave is diffusive with $g \sim 6$. Experimental details are given in Methods.

A superposition of waveguide modes is fit to the spatial distributions of the field at points on the grid at the incident and output surfaces for each of the four antenna orientations. The TM at each frequency is then expressed in terms of the waveguide modes. The TEs on the surfaces of the sample are then obtained from the SVD of the TM. This yields continuous profiles of intensity and phase at each frequency and polarization of the source and detector, such as the intensity speckle patterns in transmission for $n = 1$ and $n = 50$, shown in Figs. 1(a,b), and the corresponding phase patterns shown in Figs. 1(c,d). There are fewer speckle spots for $n = 1$ than for $n = 50$.

The number of speckle spots in the intensity pattern for a single polarization on the output surface is proportional to the number of phase singularities at which the intensity vanishes and the phase changes by $2\pi$ in a loop around a singularity[54,55]. The average number of phase singularities in polarized speckle patterns on the output surface of the TEs is plotted in Fig. 1(e) and seen to increase with $n$. This reflects the larger range of values of the transverse k-vectors for TEs with higher n, which have smaller EVs. This is seen in the plots of the weights of waveguide modes in the TEs at the sample output, $|u_{nm}|^2$, for $n = 1$ and 50 in Fig. 1(f). The waveguide modes are indexed with $m$ increasing as the group velocity falls.

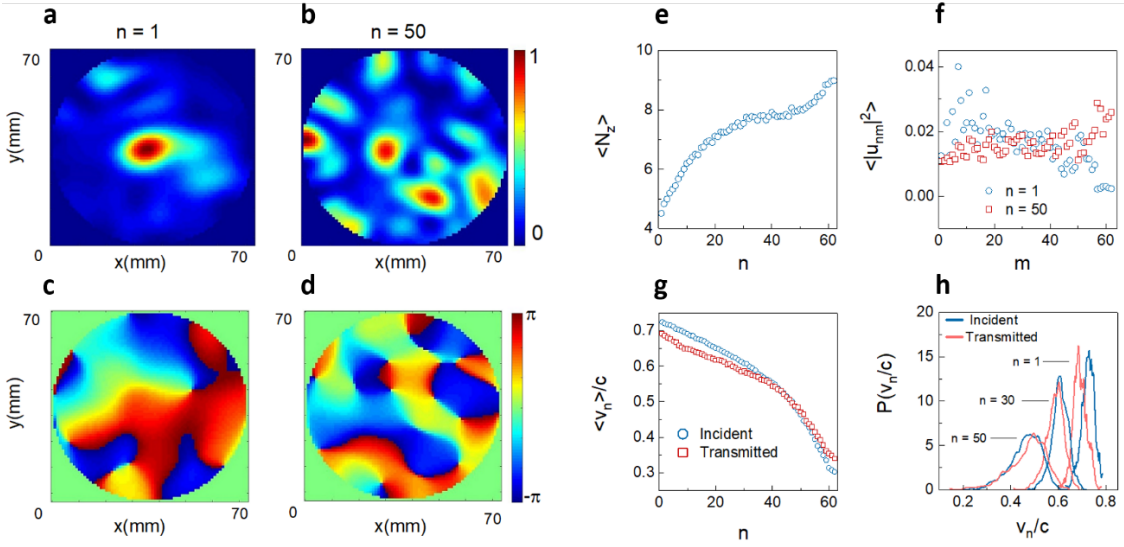

**Fig. 1: Measurements of microwave speckle and eigenchannel velocities.** (a,b) The intensity and (c,d) the phase patterns of the 1st and 50th TEs at a single frequency for a single polarization of the incident and output field in a single configuration. (e) The number of phase singularities in polarized spectra. The intensity vanishes at a phase singularity. (f) The weight of waveguide modes in the 1st and 50th TEs. (g) The EVs of the incident and transmitted waves vs. eigenchannel index, $n$. (h) The probability distribution functions (PDFs) of EVs for the incident and transmitted wave for $n = 1, 30, 50$. The differences between the input and output distributions in (g) and (h), are consistent with a small difference in orientation of the source and receiver antennas relative to the waveguide axis.

The EVs of the incident and transmitted waves for the $n^{th}$ TE, $v_{n,i}$ and $v_{n,t}$, respectively, are given by $v_{n,i} = \sum_{m=1}^{N} |v_{nm}|^2 v_{wm}$ and $v_{n,t} = \sum_{m=1}^{N} |u_{nm}|^2 v_{wm}$, where $v_{wm}$ is the group velocity of the $m^{th}$ waveguide mode. The average EVs for TEs on the input and output surfaces fall with $n$, as seen in Fig. 1(g). This reflects the increasing contributions of higher order waveguide modes with smaller axial velocities as $n$ increases. The small differences between the plots in Fig. 1(g,h) are consistent with the source antenna being more nearly perpendicular to the axis of the waveguide than the detection antenna at the output surface. Such differences are absent in the numerical simulations discussed below. Numerical simulations facilitate the study of the scaling of EVs and other propagation parameters.

**Simulations of eigenchannel velocities.** We carry out recursive Green's function simulations[56,57] of electromagnetic propagation polarized perpendicular to random 2D samples such as shown schematically in Supplementary Fig. 2. The samples of width $W$ and length $L$ are composed of square cells with sides of length $a = \lambda_0/2\pi$, where $\lambda_0$ is the free-space wavelength. The dielectric constant in each cell, $\varepsilon$, is drawn randomly from a rectangular distribution $[1 + \Delta\varepsilon, 1 - \Delta\varepsilon]$ with $\Delta\varepsilon = 0.3$. The dielectric constant is uniform in the direction perpendicular to the plane of the sample. The recursive Green's function method is discussed in Supplementary Note 1.

The scaling of the transmission eigenvalues and the EVs for $N = 8$ are shown, respectively, in Figs. 2(a) and 2(b). Whereas the $\tau_n$ fall exponentially beyond the localization length of each of the TEs and approach unity as $L \to 0$, the EVs of the transmitted wave $v_{n,t}$ saturate at distinct values as the sample length increases. For samples with large $N$, $v_{n,t}$ approaches its asymptotic value at lengths shorter than the localization length of approximately $N\ell_s$, as can be seen clearly in a sample with $N = 64$ in Supplementary Fig. 3 and Supplementary Note 2. This indicates that the different values of the EVs is a mesoscopic phenomenon unrelated to Anderson localization. The decrease in transmission for TEs with smaller EVs is consistent with the decrease in transmission with angle of incidence of an optical beam illuminating a random slab[9], as discussed in Supplementary Note 3.

The inverses of the EVs obey a sum rule because of their relationship to the average density of states (DOS) per unit angular frequency and length, which is uniform along the sample, $\rho_{\omega,L}(z) = \rho_\omega/L$. The relationship can be found by considering the sum of transmission times in all channels, given by $\tau_T = \sum_{n=1}^{N} t_n = \pi \rho_\omega$[58]. Since $\rho_\omega$ is independent of scattering strength in systems with the same average value of $\varepsilon$, it is the same as in a homogeneous sample[59]. In the present work, $\langle \varepsilon \rangle = 1$ and the transmission time for a TE in a uniform sample of unit length is $t_n = \frac{1}{v_n}$. The ensemble average local DOS (LDOS) can thus, be expressed as, $\rho_{\omega,L} = \frac{1}{\pi} \sum_{n=1}^{N} \frac{1}{v_n} \equiv \frac{N}{\pi v_+}$. As a result, $v_+$ is independent of sample length, as seen in Fig. 2(b).

The correlation of the EVs across the sample is seen in the identical PDFs of EVs with the same $n$ on the incident and output surfaces, and in the difference in EV statistics for different $n$, shown in Fig. 2(c) for $n = 1$ and 8 in a sample with $g = 1.74$. In addition, the reflected TE is proportional to the complex conjugate of the incident TE, so that the PDFs of EVs in reflection are also identical, as demonstrated in Supplementary Note 4. Thus, the average values of the EVs for the incident, reflected and transmitted waves are identical.

$$v_{n,i} = v_{n,r} = v_{n,t} \equiv v_n. \qquad (1)$$

As is conventional in discussions of the transmission eigenvalues, $\tau_n$, symbols for the EVs or other variables can refer either to the variable in a single configuration or the average over a random ensemble, depending on the context.

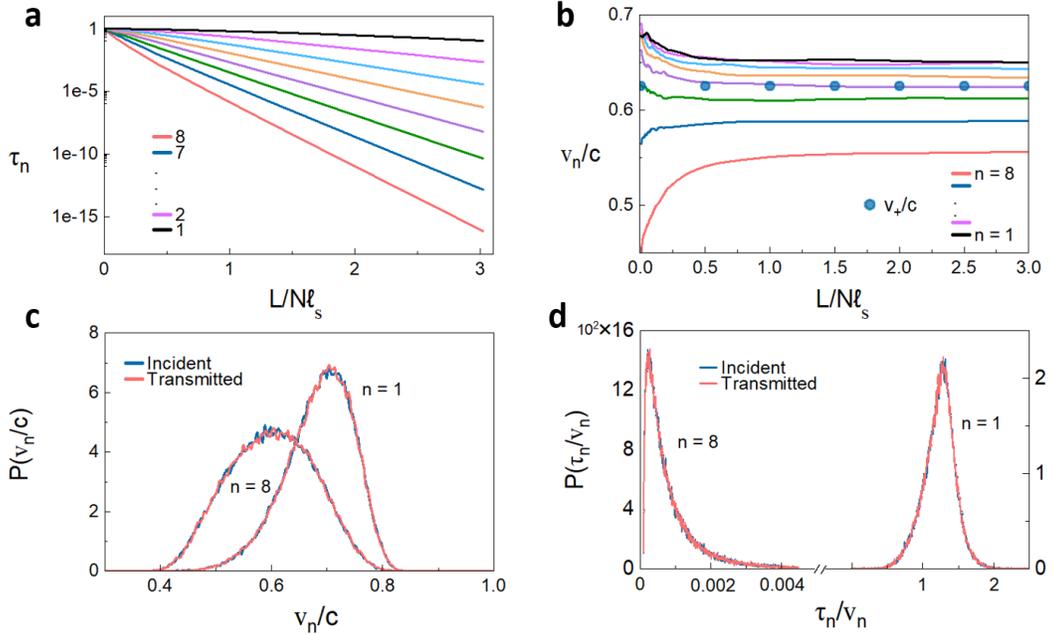

**Fig. 2:** Simulations of scaling and statistics of eigenchannel velocities. (a,b) Scaling of transmission eigenvalues and EVs in a random medium with $N = 8$. (c) PDFs of EVs of the incident and transmitted waves. For the same $n$, the PDFs overlap. (d) PDFs of $\tau_n/v_n$ for given $n$ at the input and output boundaries also overlap.

The EVs provide the link between the flux $\tau_n$ and the linear energy densities excited in TEs on the left and right boundaries of the sample, $u_n(0)$ and $u_n(L)$, respectively. For unit incident flux from the left, the linear energy densities of the incident and transmitted TEs are $u_{n,i}(0) = 1/v_{n,i}$ and $u_{n,t}(L) = \tau_n/v_{n,t}$, respectively. Since energy is conserved, the reflected flux in a transmission eigenchannel is $1-\tau_n$ and the energy density of a TE in reflection is $u_{n,r}(0) = (1-\tau_n)/v_{n,r}$. The PDFs of $\tau_n/v_{n,i}$ and $\tau_n/v_{n,t}$ are seen in Fig. 2(d) to be identical, and these are also identical to the PDFs of $\tau_n/v_{n,r}$. The average energy density in a TE at the input is found in simulations to be the sum of the averages of the energy density in the incident and reflected waves, $u_n(0) = u_{n,i}(0) + u_{n,r}(0) = \frac{1}{v_{n,i}} + \frac{(1-\tau_n)}{v_{n,r}}$. The absence of interference terms in the average energy density at the sample input surface is shown in Supplementary Note 5 to be a consequence of the proportionality of the reflected wave and the complex conjugate of the incident wave. The average energy density excited from the left is the sum over TEs, $u(z) = \sum_{n=1}^{N} u_n(z)$. This gives

$$u(0) = \sum_{n=1}^{N} \frac{2-\tau_n}{v_n}, \tag{2a}$$

$$u(L) = \sum_{n=1}^{N} \frac{\tau_n}{v_n}, \tag{2b}$$

on the left and right sides of a dissipationless sample.

**The diffusion coefficient.** When the energy density within the sample excited from the left falls linearly, it is possible to define a diffusion coefficient via Fick's first law as the ratio of the flux and the magnitude of the gradient of the energy density, $D = -g/\frac{du}{dz}$, with $\frac{du}{dz} = \frac{u(L)-u(0)}{L}$, giving

$$D = \frac{gL}{u(0)-u(L)}. \qquad (3)$$

With the energy densities at the sample boundaries given in Eq. (2), this gives

$$D = \frac{gL}{2\left[\sum_{n=1}^{N}\frac{1}{v_n} - \sum_{n=1}^{N}\frac{\tau_n}{v_n}\right]}. \qquad (4a)$$

The diffusion coefficient may be expressed in terms of $v_+$ and an effective transmission velocity, $v_T$, defined via the relation, $u(L) = \sum_{n=1}^{N}\frac{\tau_n}{v_n} \equiv \frac{g}{v_T}$, as

$$D = \frac{gL}{2\left[\frac{N}{v_+} - \frac{g}{v_T}\right]}. \qquad (4b)$$

The diffusion coefficient can be defined as in Eqs. (3) and (4) even for $g = 1.74$, which is close to the crossover to Anderson localization, since $du(z)/dz$ is essentially constant throughout the sample, as seen in Fig. 3. $u(z)$ is normalized in the figure by its spatial average, $\langle u(z)\rangle_z = \frac{N}{v_+}$, which is calculated in Supplementary Note 6. The figure also shows the normalized energy density in a single configuration. The configuration average of $u(z)v_+/N$ falls linearly in the sample and extrapolates to zero a distance $z_b$ to the right of the sample and to 2 at a distance $z_b$ to the left of the sample. Since the triangles in Fig. 3 with bases of $L + 2z_b$ and height 2 and with base $z_b$ and height $u(L)v_+/N$ are similar, $\frac{2}{L+2z_b} = \frac{u(L)v_+/N}{z_b}$ [60], and

$$u(L) = \frac{2Nz_b/v_+}{L+2z_b}. \qquad (5)$$

Since $g = u(L)v_T$, this gives

$$g = \frac{2Nz_b}{L+2z_b}\frac{v_T}{v_+}. \qquad (6)$$

Substituting Eq. (6) into Eq. 4(b) gives

$$D = z_b v_T, \qquad (7)$$

as shown in Supplementary Note 7. Unlike, the Boltzmann diffusion coefficient, $D_B$, the diffusion coefficient, $D$, is expressed here in terms of the nature of the wave near the boundaries rather than in the bulk of the medium. As seen below, $D$ and its factors in Eq. (7) depend upon the dimensions of the sample even in the diffusive limit.

The variation of $D$ with length and width for samples with $\Delta\varepsilon = 0.3$ and $N = 8, 16, 32,$ and 64 channels is shown in Figs. 4a and 4b, respectively. To place the results in the context of particle diffusion, $D$ is normalized by $\widetilde{D_0} = \frac{1}{d}c\ell_s$, where $d = 2$ is the dimensionality, $c$ is the speed of the wave in a medium with $\varepsilon = 1$, and $\ell_s$ is the scattering mean free path in which the wave loses coherence. $\ell_s$ is determined from the identical decay times of the coherent flux for all waveguide modes[60], as discussed in Supplementary Note 8 and shown in Supplementary Fig. 4. $\widetilde{D_0}$ approximates the bare particle diffusion coefficient, $D_0 = \frac{1}{2}v_E\ell$[2,3,61]. Here, $v_E$ is the energy transport velocity[61], which may be lower than the phase velocity in the medium as a result of resonances with scattering elements. However, since the sides of the scattering elements of the sample are much shorter than the wavelength, $a = \lambda_0/2\pi$, these elements are far from resonance. Scattering should therefore be nearly isotropic, with $v_E \sim c$, $\ell_s \sim \ell$, and $\widetilde{D_0} = \frac{1}{2}c\ell_s \sim D_B$.

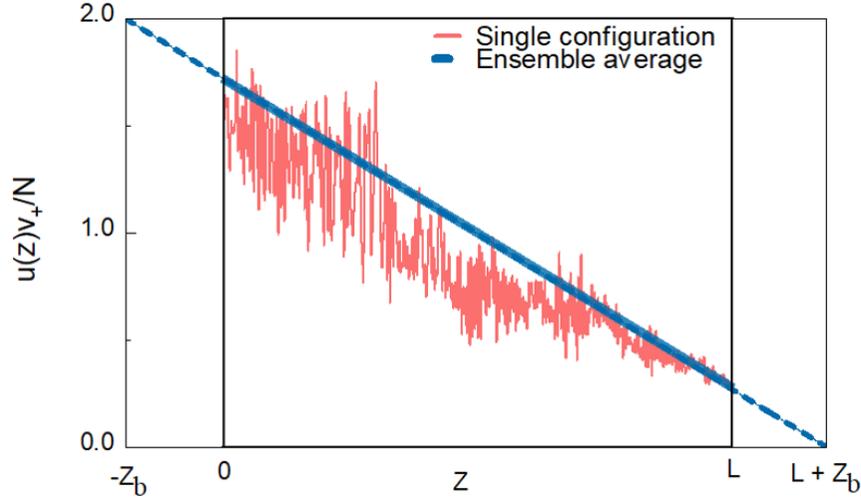

**Fig. 3:** Energy density inside the random medium. Profiles of the energy within a medium excited from the left with unit flux in all channels in a single configuration (red curve) and averaged over 2,000 configurations (blue curve) of a medium with $N = 8$ and $g = 1.74$. The average energy density falls linearly and extrapolates to zero at a distance $z_b$ beyond the sample.

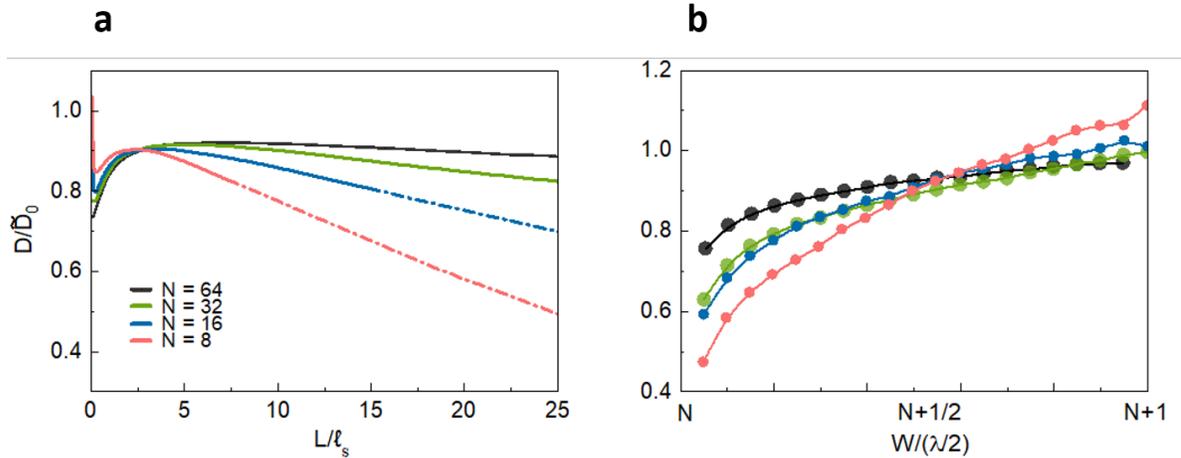

**Fig. 4:** Scaling of the diffusion coefficient. (a) $D(W, L)$ for $W = (N + \frac{1}{2})\lambda_0/2$ is plotted for different values of $W/(\lambda_0/2)$. The diffusion coefficient may be defined according to Fick's first law as long as the derivative of energy density within the sample is nearly constant, which is the case down to values of $g$ close to unity. The diffusion coefficient, as given by Eqs. (3) and (4), is plotted as dashed curves, since $u(z)$ then no longer falls linearly. (b) The variation of the diffusion coefficient with wavelength for each $N$ is plotted for sample lengths at which the diffusion coefficient in (a) is at its maximum.

The variation of $D$ with length for samples with width in the centre of the range for each channel, $W \sim \left(N + \frac{1}{2}\right)(\lambda_0/2)$ is shown in Fig. 4(a). $D(L)$ reaches a peak after several scattering mean free paths and then falls due to weak localization[12,62]. The decay is more rapid for samples with smaller $N$ since the crossover to localization at $g \sim 1$ at length $\xi \sim N\ell \sim N\ell_s$, is reached at shorter lengths. The value of $D$ found from the decay of pulsed transmission following the peak in transmission is the same as obtained in steady-state simulations, as shown in Supplementary Note 9 and Supplementary Fig. 5.

The diffusion coefficient also changes as the ratio of the sample width and wavelength is varied by changing the wavelength in a sample with the same number of square scattering elements with sides of length $\lambda/2\pi$. The variation of $D$ over the range of width of $[N, N+1]\lambda_0/2$ for the same values of $N$ and $\lambda_0$ as in Fig. 4(a) at values of $L/\ell_s$ at which $D/\widetilde{D_0}$ reaches its peak is shown in Fig. 4(b). $D$ varies because the shapes of the propagating modes changes with sample width. The fractional variation of $D$ over the range of wavelength for a given $N$ decreases as $N$ increases but is still appreciable for $N = 64$.

Comparing Eq. (7) to the classical expression for the diffusion coefficient in $d$ dimensions, $D_B = \frac{1}{d} v_E \ell$, gives, $\ell = d z_b v_T / v_E$. Below, we find that $D$, and its factors $z_b$ and $v_T$ vary with the dimensions of the sample so that $\ell$ that appears in the Boltzmann is not an intensive parameter, as is often assumed. In the extreme diffusive limit, however, in which $N \gg g \gg 1$, $D$ and its factors $z_b$ and $v_T$ would be expected to approach the classical particle limit, in accord with the correspondence principle.

The variation of $z_b$ and $v_T$ with length for samples with $W = (N + \frac{1}{2})\frac{\lambda_0}{2}$ for $N = 8, 16, 32, 64$ are shown in Figs. 5(a) and 5(b). Since $v_T$ varies little for $L/\ell_s > 2$, the fall in $D/\widetilde{D_0}$ with $L$ after the crossover from ballistic to diffusive propagation primarily reflects the variation of $z_b/\ell_s$. The values of $z_b$ in Fig. 5(a) are obtained by solving Eq. (6). This gives

$$z_b = g L/2 \left[N \frac{v_T}{v_+} - g\right]. \quad (8)$$

The values of $z_b/\ell_s$ in the centre of the wavelength range for a given $N$ are close for all values of $N$, as seen in Fig. 5(c). For $N = 64$, $z_b/\ell_s = 0.711$. This is close to the value obtained from the solution of the Milne problem of 0.7104 for equilibrium radiative transfer near the surface of a half space due to a remote source within the medium[2,5]. However, $z_b/\ell_s$ varies with both sample length and width. The variation with sample width of $z_b/\ell_s$, $v_T/c$, $z_b$, and $\ell_s$ is shown in Figs. 5(c-f) for sample lengths at which $D/\widetilde{D_0}$ reaches its peak value for each value of $N$ at sample widths $W = (N + \frac{1}{2})\frac{\lambda_0}{2}$. The variation of these parameters with width for $N = 8$ and 64 and a large range of lengths is shown in Supplementary Fig. 6 and Supplementary Note 10.

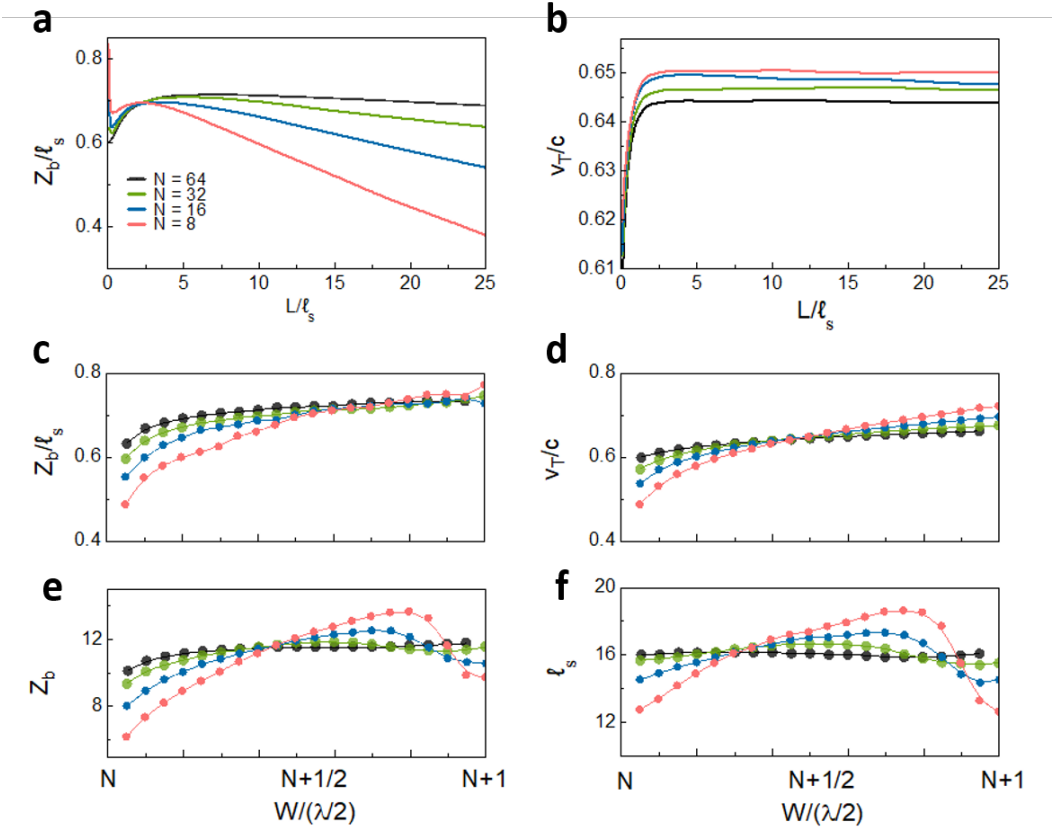

**Fig. 5:** Scaling of the factors of the diffusion coefficient. (a,b) The scaling of $z_b/\ell_s$ and $v_T/c$ with length for various values of $N$. $z_b$ falls linearly with length while $v_T$ become nearly independent of $L$ after two scattering lengths. (c-f) The parameters $\frac{z_b}{\ell_s}$ (c), $\frac{v_T}{c}$ (d), $z_b$ (e), and $\ell_s$ (f) all vary with sample width. The variation of $\ell_s$ with $W$ demonstrate that the scattering process is nonlocal.

Closer to the Anderson localization threshold, the diffusion model breaks down with $u(z)$ falling faster in the centre of the sample than at the boundaries, as seen in Figs. 6(a), so that it is not possible to define a diffusion coefficient. For $g = 1.078$, $u(z)$ falls 10% faster at the centre than at the edges of the sample, as is seen in Supplementary Fig. 7 and discussed in Supplementary Note 11. Transport can then be described in terms of a position-dependent diffusion coefficient, $D(z) \equiv -g/\frac{du(z)}{dz}$ [63,64], which dips in the middle of the sample because of the larger probability of trajectory crossing themselves there than near the sample boundaries.

The energy density profiles in Figs. 3, 6a, and 12 are the sums of the energy density of TEs $u_n(z)$, such as the profiles shown in Figs. 6(b-d), with values at the boundaries that depend upon $\tau_n/v_n$, as in Eq. (2). In addition to the larger magnitude of the slope of $u(z)$ in the centre relative to that at the boundaries of the sample as $g$ decreases, as seen in Fig, 6(a), there is an inversion in the ranking of energy excited in the sample vs. eigenchannel index, $n$, in the crossover from ballistic to diffusive propagation. In translucent samples, all of the $\tau_n$ are close to unity and the energy density throughout the sample is dominated by the factor $1/v_n$, and so the energy within the sample is larger for smaller transmission eigenvalues, as seen in Fig. 6(b). This trend is reversed in longer samples, as seen in Figs. 6(c,d). For $L > \ell_s$, the variation of $\tau_n/v_n$ with $n$ is predominantly due to the sharp variation of $\tau_n$ with $n$, and the energy excited in the sample decreases with $n$, as seen in Figs. 2(a) and 2(b).

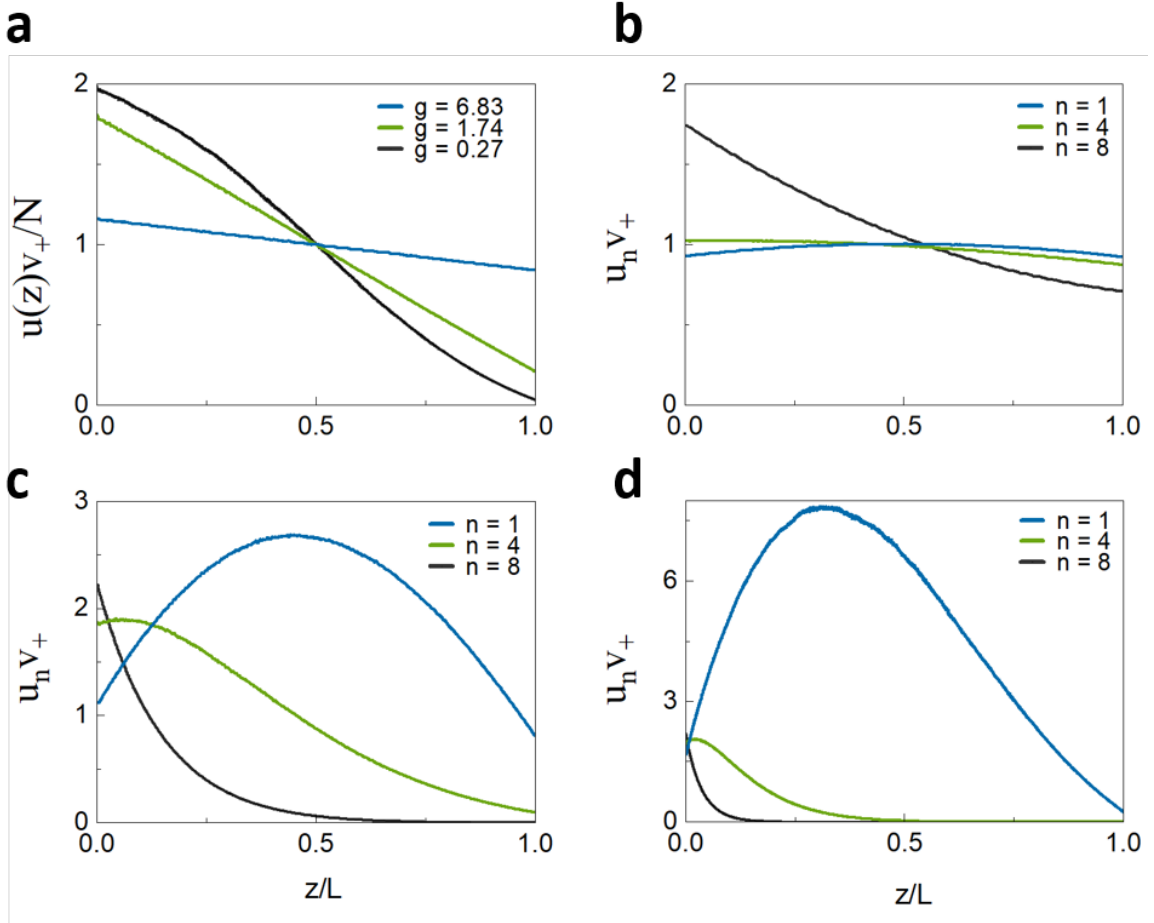

**Fig. 6:** Profiles of excitation inside the medium. (a) The sum over all TEs of the average energy density excited within the medium from the left for ballistic, diffusive, and localized waves. The energy density falls linearly throughout the sample for ballistic and diffusive waves but falls more rapidly near the centre for localized waves. (b-d) The energy excited in TEs increases with $n$ for ballistic waves but decreases with $n$ for diffusive and localized waves.

**The Thouless Conductance.** The scaling theory of localization[25,65], according to which the variation of $g$ with the dimensions of the sample depends only upon $g$, is built upon the relationship between static and dynamic aspects of wave transport. The classical geometric model of the scaling of conductance in the diffusive limit, $g = \frac{A\sigma}{L\left(\frac{e^2}{h}\right)}$, may be expressed in terms of dynamic parameters via the Einstein relation for the conductivity, $\sigma = e^2 D \rho_{E,A,L}$ in terms of the local parameters of the diffusion coefficient and the LDOS per unit energy and volume, $\rho_{E,A,L} = \rho_E/AL$[25]. This gives $g = \frac{hD\rho_E}{L^2}$. The classical wave analogue of this relation is $g = \frac{2\pi D \rho_\omega}{L^2}$.

Using the results in the previous section, we obtain a relationship between $g$, $D$, and $\rho_\omega$ by substituting Eq. (7) into Eqs. (6) and utilizing the relation, $\rho_\omega = \frac{NL}{\pi v_+}$, to give

$$g = \frac{2\pi D \rho_\omega}{L(L+2z_b)}. \tag{9}$$

Equation (9) is valid as long as $u(z)$ falls linearly within the sample, which is the case even quite close to the localization threshold.

Following Thouless, the right-hand side of Eq. (9) may be expressed in terms of the degree of spectral overlap of the quasi-normal modes, or resonances, of the open medium, which is the ratio of the modal linewidth and the spacing between modes, $\delta = \delta\omega/\Delta\omega$ and is known as the Thouless number or the Thouless conductance, $\delta \equiv g_{Th}$. The linewidth of quasi-normal modes in a diffusive sample, $\delta\omega = \frac{\pi^2 D}{(L+2z_b)^2}$, is equal to the decay rate of stored energy following pulsed excitation and also to the decay rate of the lowest mode of the diffusion equation[20], while the DOS is equal to the inverse of the average spacing between modes, $\rho_\omega = \frac{1}{\Delta\omega}$. Even as $D(W,L)$ is renormalized by weak localization, its value as determined from Fick's first law is still identical to that obtained from the decay rate of energy following pulsed excitation, as seen in Fig. 11 in Appendix I. Equation (9) can thus be expressed as $g = \frac{2(L+2z_b)}{\pi L} \frac{\delta\omega}{\Delta\omega} \equiv \frac{2(L+2z_b)}{\pi L} \delta$.

The Thouless conductance may be compared to the degree of crossing of wave trajectories, which is the ratio of the average time to for waves to traverse the sample, $\tau_{Th}$, to the time to visit each coherence volume in the sample, $\tau_T$, $\frac{\tau_{Th}}{\tau_T}$. Since $\tau_T = \pi \rho_\omega$ and, $\tau_{Th} = (L+2z_b)^2/D$[22], and $\Delta\omega$ and $\delta\omega$ are as given above, this gives, $\delta = \pi \frac{\tau_{Th}}{\tau_T}$, and Eq. (9) can be written as, $g = \frac{2(L+2z_b)}{L} \frac{\tau_T}{\tau_{Th}}$. Thus, whether because of low modal overlap, $\delta$, or a large fraction of coherence lengths along the trajectory that are crossed by a trajectory, $\frac{\tau_{Th}}{\tau_T}$, $g$ is suppressed below the predictions of diffusion theory as $g$ falls towards and below unity[25,65,16,11,66,67]. Mesoscopic fluctuations are then greatly enhanced over the level predicted by Gaussian field statistics[13,16,66,68,69], transmission spectra are sharply peaked[70–72], and energy density falls exponentially from the point at which it is injected[73–75]. Since $1/g$ expresses the degree of departure from diffusion theory, the longitudinal scaling of the conductance should depend only upon $g$ itself[25,65]. However, the oscillations in scattering parameters shows that the scaling in the transverse dimension requires the introduction of a second dimensionless parameter, $W/\lambda$. This parameter determines the propagating channels which are coupled by the disorder of the medium.

**Discussion**

The absence of randomization of the velocity of transmission eigenchannels in mesoscopic samples ushers in the EVs, $v_n$, as a new set of parameters of the TM alongside the transmission eigenvalues, $\tau_n$. The average values and statistics of different EVs are different, but they are identical on the input and output surfaces of the sample for the same TE. This makes it possible to utilize the TM to explore dynamic as well as steady-state propagation,

The energy densities of different TEs on the output surface may be expressed using the sets of $\tau_n$ and $v_n$ since $u_n(L) = \tau_n/v_n$. Thus, the gradient of energy density within the sample and the diffusion coefficient can be found. With this diffusion coefficient, transport in steady-state and in the time domain can be described via Fick's first and second laws, respectively. The value of the diffusion coefficient found from simulations of pulsed transmission is the same as found from the TM.

We have found that the diffusion coefficient, $D$, the boundary extrapolation length, $z_b$, and the effective longitudinal velocity in transmission, $v_T$, may vary appreciably over the range of wavelength or sample width in which the number of channels exciting the sample increase by unity. Since all these quantities dip near the crossover to a new channel as the sample width or wavelength is changed, even in the diffusive regime, these parameters are global rather than local and their variation with the transverse dimensions of the sample is unrelated to wave localization. The degree of modulation of these parameters with $W/(\lambda/2)$ falls as the size of the sample increases in harmony with the correspondence principle of quantum mechanics.

The nonlocality of propagation is seen in the expression for the diffusion coefficient as a product of two factors which reflect propagation at the sample boundaries rather than in its interior, $D = z_b v_T$. This expression for $D$ differs from the Boltzmann diffusion coefficient, $D_B = \frac{1}{d} v \ell$, expressed in terms of local parameters, which are generally assumed to be independent of the sample's dimensions.

We have also found in simulations that even the scattering mean free path, $\ell_s$, which might be expected to represent local scattering, dips near the crossover to a new channel. The scattering elements in the simulations are small relative to the wavelength, so that $\ell \sim \ell_s$. We find that the ratio $z_b/\ell_s \sim z_b/\ell$, which has been assumed to have a constant value of 0.7104 obtained in the solution of the Milne problem in an unbounded sample[2], also varies with sample width. When the sample width is close to the centre of the range of widths for a given value of $N$, however, this ratio is close to the particle diffusion value found in the solution of the Milne problem.

Systematic variations of the conductance and transmittance as the number of propagating channels changes also arise in ballistic and diffusive samples. Stepwise increases in the conductance are measured in ballistic heterojunctions as the width is changed by a gate voltage[76] and also in the optical transmittance of diffuse light through an adjustable aperture[77]. Simulations of conductance in diffusive quasi-1D samples were carried for the Anderson model of a tight-binding Hamiltonian with diagonal disorder. Instead of steps, dips were found in the conductance near the threshold to each new channel[78,79]. The origin of the dips will be shown in future work to emerge from the correlation within and between the sets of $\tau_n$ and $v_n$.

The modulation of scattering with the number of propagation channels is reminiscent of the Wigner cusp anomaly in the nuclear scattering cross section that arises when new scattering channels open up as the energy of incident particles increases[82–84]. The enhanced variation of scattering around the crossover to a new channel can be the basis for enhanced sensitivity of classical waves to variations in the sample dimensions[85]. Present research is exploring the extreme sensitivity of EVs to changes in the sample dimensions at the crossover to a new channel. Pressing questions that emerge from this work are the relationships between the EVs and the transmission eigenvalues and the profiles of EVs within the sample.

**Methods**
**Microwave Measurements**
In- and out-of-phase spectra of field transmission coefficients between source and receiver antennas on opposite sides of the sample are shown in Fig. 7(a). Such spectra are obtained for each of four pairs of polarizations on the input and output surfaces for each pair of locations of the 4-mm-long source and receiving antennas on a square grid with 9-mm spacing on the sample's surfaces. A schematic of the experimental setup is shown in Fig. 7(b). A copper sample tube of inner-diameter 7.3 cm is filled to a length of 23 cm with 0.95-cm-diameter alumina spheres of refractive index 3.14 embedded in the centres of Styrofoam shells to yield an alumina volume fraction of 0.07. An ensemble of 23 random sample configurations is created by briefly rotating the tube about its axis and vibrating the sample. The sample is vibrated to allow the sample to settle so that it is static over the 24-hour period required to measure the spectrum of the TM. The antennas are formed by stripping the outer conductor and bending the central conductor of a microwave cable by approximately 90°. The bent central conductors of the source and detector are brought close to the input and output surfaces and lie parallel to the respective surfaces.

The electric fields launched into and emerging from the sample are polarized along the direction of the respective wire antennas. The fluxes on the incident and output surfaces are found by expressing the field amplitudes of the incoming and outgoing waves as superpositions of waveguide modes.

**Data Availability**
The datasets generated during and/or analysed during the current study are available from the corresponding author on reasonable request.
**Data Availability**
The simulation codes used in the current study are available from the corresponding author on reasonable request.

———

**Acknowledgements**

We thank Krishna Joshi for plotting Figs. 1(a-d) and for discussions and thank Israel Kurtz for discussions. This work is supported by the National Science Foundation (US) under EAGER Award No. 2022629 and NSF-BSF Award No. 2211646.

**Author contributions**

A.Z.G directed the project, calculated propagation parameters and wrote the manuscript with input from all authors. Z.S. carried out the measurements. Y.H. analysed the experimental data, carried out and analysed numerical simulations, and calculated the relationship between the incident and reflected TEs. A.M. carried out and analysed numerical simulations.

**Competing Interests**

The authors declare no competing interests

**Additional information**

**Supplementary information** The online version contains supplementary information

**Correspondence** and requests for materials should be addressed to Azriel Z. Genack at agenack@qc.cuny.edu

**Supplementary Information for "Transmission-eigenchannel velocity and diffusion"**

Azriel Z. Genack[1,2*], Yiming Huang[1,2,3], Asher Maor[1,2,4] and Zhou Shi[1,2,5]
[1]Department of Physics, Queens College of the City University of New York, Flushing, New York 11367, USA
[2]Physics Program, The Graduate Center of the City University of New York, New York New York, 10016, USA
[3]Jinhua No.1 High School, Zhejiang, 321000, China
[4]Kent Optronics, Inc., Hopewell Junction, New York 12533, USA
[5]OFS Labs, 19 School House Road, Somerset, New Jersey 08873, USA

**Contents**
**Supplementary Figures**
1. Microwave measurements and experimental setup
2. Model of random sample
3. Scaling of eigenchannel velocities for $N = 64$
4. Determination of the scattering mean free path, $\ell_s$
5. Simulation of pulsed transmission
6. Scaling of diffusion coefficient and its factors
7. Departure from linearity of $u(z)$ near the localization threshold

**Supplementary Notes**
1. Recursive Green's function simulations
2. Scaling of eigenchannel velocities for $N = 64$
3. Comparison of microwave and optical measurements of transmission.
4. Relationship between incident and reflected transmission eigenchannels
5. Energy density at the input surface
6. Normalization of energy density profile by its spatial average
7. Expression for nonlocal diffusion coefficient
8. Determining the scattering mean free path
9. Pulse propagation
10. Scaling of diffusion coefficient and its factors
11. Breakdown of diffusion in energy density profile

**Supplementary References**

# Supplementary Figures
## 1. Microwave measurements and setup

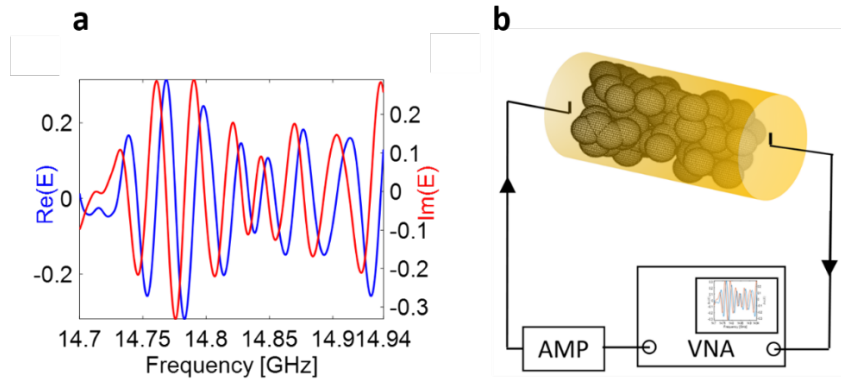

**Supplementary Fig. 1** | Microwave measurements and setup. (a) Microwave spectra of in- and out-of-phase spectra of field transmission coefficients between source and receiver antennas on opposite sides of the sample for a single polarization pair on the incident and output surfaces of the sample. **b**, Schematic of experimental setup showing the vector network analyzer, amplifier of the source, and bent antennas.

## 2. Model of random sample

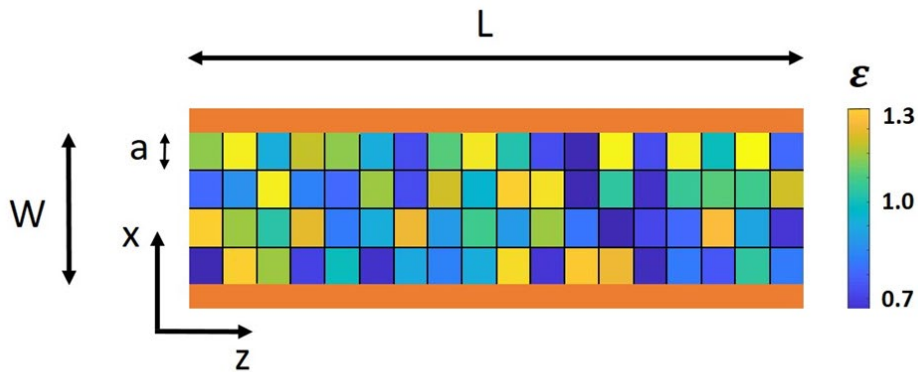

**Supplementary Fig. 2** | Model of random sample. Model of random sample. The random sample is modelled by a lattice of square elements with sides of length $a = \lambda_0/2\pi$ and random values of the dielectric constant.

## 3. Scaling of eigenchannel velocities for $N = 64$

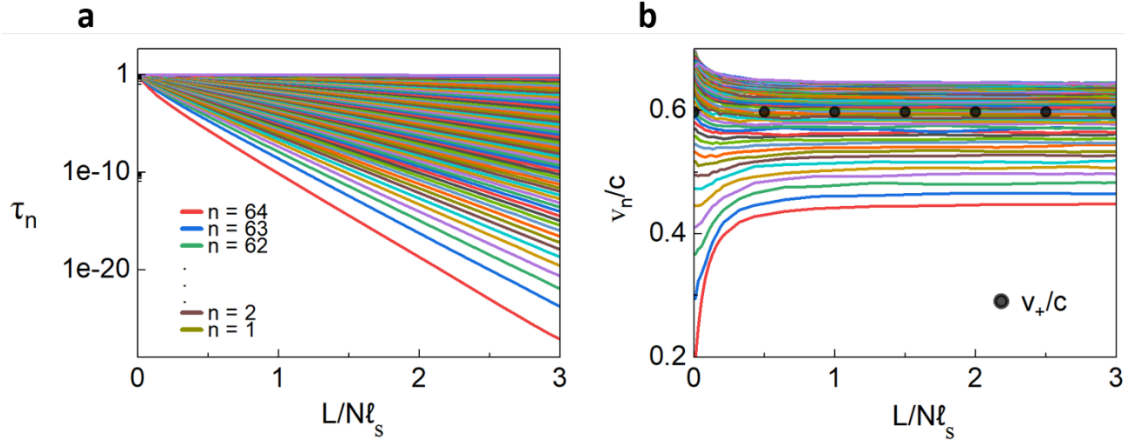

**Supplementary Fig. 3** | Scaling of eigenchannel velocities for $N = 64$. (a) The transmission eigenvalues fall exponentially with sample length, while (b) the EVs approach distinct values asymptotically on length scales of a few times $\ell_s$. For large $N$, this occurs for lengths shorter than the localization length of approximately $N\ell_s$.

## 4. Determination of the scattering mean free path, $\ell_s$

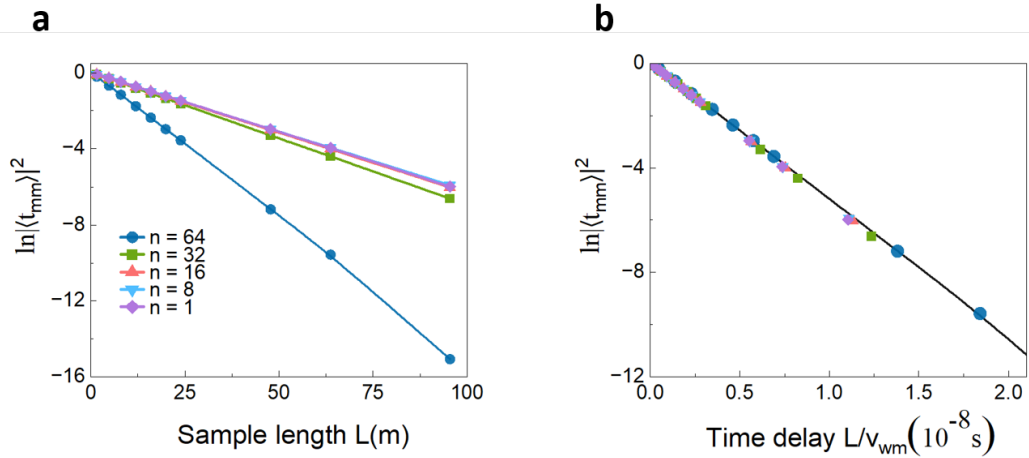

**Supplementary Fig. 4** | Determination of the scattering mean free path, $\ell_s$. (a) Scaling of the coherent flux of waveguide modes, $|\langle t_{mm}\rangle|^2$ in a sample with $N = 64$. (b) The decay of $|\langle t_{mm}\rangle|^2$ with coherent time delay $L/v_{wm}$, where $v_{wm}$ is the group delay of the $m^{\text{th}}$ waveguide mode. The decay time is the same for all waveguide modes.

## 5. Simulation of pulsed transmission

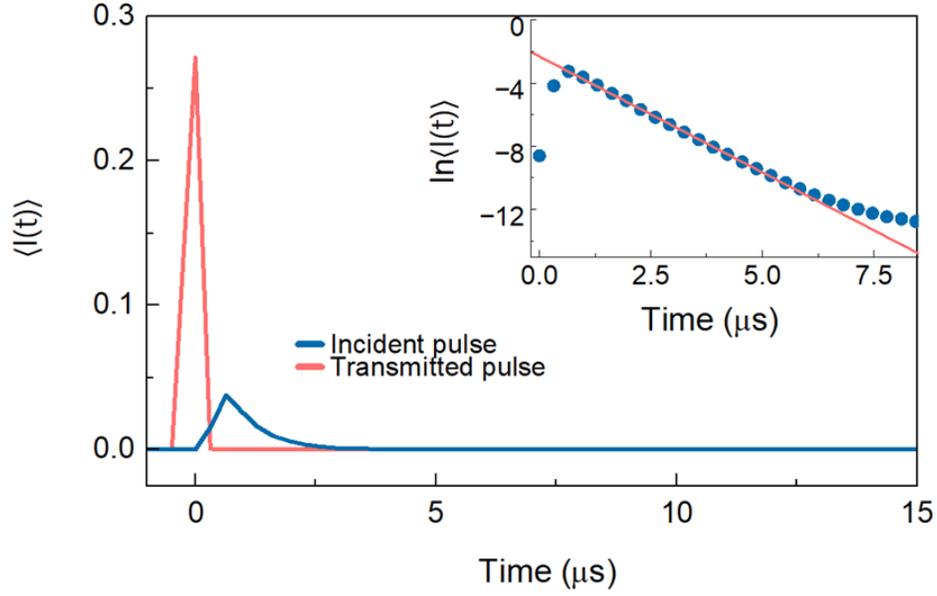

**Supplementary Fig. 5 |** Simulation of pulsed transmission. Pulsed transmission in sample with $N = 64$, $L = 7\ell_s$, and $\Delta\varepsilon = 0.3$. The inset shows that the decay of $\ln\langle I(t)\rangle$, with a rate that is proportional to the diffusion coefficient.

## 6. Scaling of diffusion coefficient and its factors

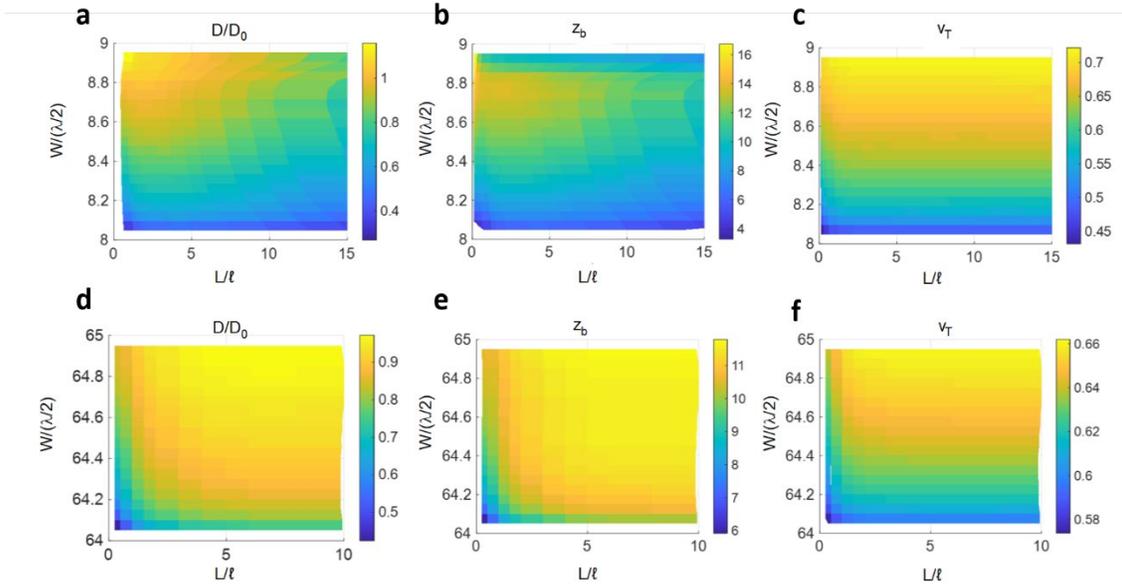

**Supplementary Fig. 6 |** Scaling of normalized diffusion coefficient and its factors. Scaling of $D/\widetilde{D_0}$, $z_b$ and $v_T$ with width and length are shown in the top row (a-c) for $N = 8$ and in the bottom row (d-f) for $N = 64$.

## 7. Departure from linearity of $u(z)$ near the localization threshold

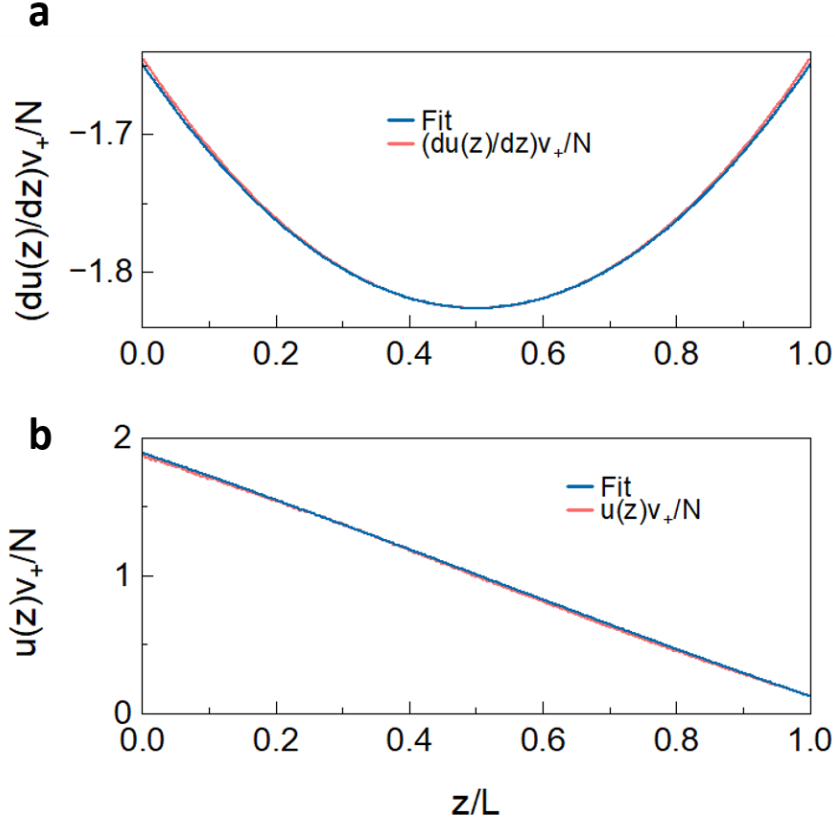

**Supplementary Fig. 7 |** Departure from linearity of $u(z)$ near the localization threshold.

**Supplementary Notes**
**1. Recursive Green's function simulations**
Simulations of electromagnetic wave propagation are carried out on model rectangular samples shown schematically in Supplementary Fig. 2. Samples of length $L$ and width $W$ are comprised of a lattice of squares with sides $a = \lambda/2\pi$ and random values of the dielectric constant, $\varepsilon$, drawn from a rectangular distribution $[1 - \Delta\varepsilon, 1 + \Delta\varepsilon]$ with $\Delta\varepsilon = 0.3$. The sample is open on the left and right and bounded on top and bottom by perfect reflectors. Thus, field excited from the open boundaries of the sample may be expressed in terms of the waveguide modes of the empty waveguide.

We employ the recursive Green's function method[1–3] to find the field in the $k^{\text{th}}$ column for a source on the left hand side of the sample. The sample is divided into $K$ columns of width $a$, labeled 1, 2, …, k, … K. To find the field in each column of the sample, we first consider the field in an isolated column with random dielectric constant in each element within an otherwise empty waveguide. The field in the isolated $k^{\text{th}}$ column is found by solving the homogeneous wave equation $(E_{kk} - H_{kk})\Psi = 0$, where $\Psi$ is the electric field, $H_{kk} = \nabla^2 + k^2$, and $E_{kk}$ is the eigenvalue of the wave equation. We refer to $H_{kk}$ as a Hamiltonian because the wave equation is discretized in analogy to the tight-binding Hamiltonian used in electronic systems[1].

We seek to solve the equation $(E_{kk} - H_{kk}(\mathbf{r},\mathbf{r}'))G_{kk}(\mathbf{r},\mathbf{r}') = \delta(\mathbf{r}-\mathbf{r}')$ for a point source. We are interested in the causal solution for the Green's function, $G_{kk} = \lim_{\eta \to 0^+}(E_{kk} + i\eta - H_{kk})^{-1}$. The total Hamiltonian of the system is:

$$\begin{pmatrix} H_L & H_{1L} & 0 & 0 & 0 & 0 \\ H_{L1} & H_{11} & H_{12} & 0 & 0 & 0 \\ 0 & H_{21} & H_{22} & \ddots & 0 & 0 \\ 0 & 0 & \ddots & \ddots & H_{K-1,K} & 0 \\ 0 & 0 & 0 & H_{K,K-1} & H_{KK} & H_{KR} \\ 0 & 0 & 0 & 0 & H_{RK} & H_R \end{pmatrix}$$

Here, $H_L$ is the Hamiltonian of the left lead, $H_R$ is the Hamiltonian of the right lead, $H_{kk}$ is the Hamiltonian of the $k^{th}$ column, and $H_{kl}$ is the interaction term between the neighboring $k^{th}$ and $l^{th}$ columns. We begin with the surface Green's function of the left lead and proceed from left to right, connecting each column to the subsystem to its left.

The coupling between a $k^{th}$ column and the subsystem to its left is expressed via the Dyson equation, $G = G_0 + G_0 V G$. $G$ is the Green's functions of the connected columns from 1 to $k$. This is the matrix of solutions of the Hamiltonian submatrix from 1 to $k$. $G_0$ is the matrix of Green's functions of the connected columns from 1 to $k$-1 and the isolated column $k$, for a subsystem with the Hamiltonian

$$H_0 = \begin{pmatrix} H_L & H_{1L} & 0 & 0 & 0 \\ H_{L1} & H_{11} & H_{12} & 0 & 0 \\ 0 & H_{21} & H_{22} & \ddots & 0 \\ 0 & 0 & \ddots & \ddots & 0 \\ 0 & 0 & 0 & 0 & H_{kk} \end{pmatrix}$$

and $V$ is the interaction terms between the two, given by

$$V = \begin{pmatrix} 0 & 0 & 0 & 0 & 0 \\ 0 & 0 & 0 & 0 & 0 \\ 0 & 0 & 0 & \ddots & 0 \\ 0 & 0 & \ddots & \ddots & H_{k-1,k} \\ 0 & 0 & 0 & H_{k,k-1} & 0 \end{pmatrix}$$

This gives the recursion equation

$$G^L_{k+1,1} = G^L_{k+1,k+1} V_{k+1,k} G^L_{k,1}, \tag{10}$$

where the superscript $L$ indicates the connected subsystem starting at the left side, $G_{kl}$ is the Green's function of the connected subsystem at column $k$ beginning from column $l$, and $G^L_{k+1,k+1} = [1 - G_{k+1,k+1} V_{k+1,k} G^L_{kk} V_{k,k+1}]^{-1} G_{k+1,k+1}$. A similar recursion relation is found by iterating from right to left. The left-to-right and right-to-left parts are then combined to obtain the result, $G_{k1} = [1 - G^L_{kk} V_{k,k+1} G^R_{k+1,k+1} V_{k+1,k}]^{-1} G^L_{k1}$, which allows us to calculate the Green's functions for all columns from 1 to K. The Green's functions are used to calculate the energy density, TM, transmission eigenvalues and EVs, and the energy density in each TE.

## 2. Scaling of eigenchannel velocities for $N = 64$

In contrast to the exponential decay of transmission eigenvalues with length in samples with $N = 8$, the EVs asymptotically approach a constant value with increasing length, as seen in Fig. 2(a,b). Most EVs approach an asymptotic value at lengths shorter than the localization length of approximately $N\ell_s$, but the EV of the transmission eigenchannels with the lowest transmission approaches its asymptotic value at a length which is comparable to $N\ell_s$. To check whether the approach of EVs to their asymptotic values is related to the onset of multiple scattering or of localization, we carried out simulations in samples with the same disorder of $\Delta\varepsilon = 0.3$ but with $N = 64$. The localization length is then much longer than the scattering mean free path. The results in Supplementary Fig. 3 show that the EV of low transmission eigenchannels close approach their asymptotic values for lengths decidedly shorter than $N\ell_s$. This indicates that the evolution of values of EVs with length is related to $\ell_s$ and not $N\ell_s$. Thus, the EV do not change in this sample in the crossover to Anderson localization and their distinct values are not related to Anderson localization.

### 3. Comparison of microwave and optical measurements of transmission

The nature of transmission and energy density within a random medium depends upon the excitation. When the TM can be measured, it is possible to excite the medium with the sum of all incident waveguide modes, as is the case in the present paper. This occurs naturally in measurements of electrical conductance. it is possible to excite Lower values of EVs for channels with smaller $\tau_n$ is consistent with measurements of a drop in transmission of a laser beam directed at a random dielectric slab with increasing angle of incidence, which corresponds to a reduced normal component of the velocity of the light [26]. The transmission through the slab is $T(\theta) = \frac{z_p \cos\theta' + z_b}{L + 2z_b}$, where $z_p$ is the distance traveled by the beam before the direction light is randomized, and $\theta'$ is the angle of refraction within the sample.

The derivative of energy density is constant within a random medium and $z_b$ is clearly defined when the flux is the same in all incident channels. More generally, as for the case of a laser beam incident on a random slab, the energy density will be a weighted superposition of the eigenchannels and will produce a different spatial profile of energy density, even deep within a random medium.

### 4. Relationship between incident and reflected transmission eigenchannels

For a dissipationless quasi-1D medium with reciprocity, the scattering matrix can be expressed as $S = \begin{bmatrix} r & t' \\ t & r' \end{bmatrix}$, where $r$ and $t$ are the reflection and transmission matrices for a wave incident from the left and the corresponding matrices for excitation from the right are primed.

Reciprocity implies $S^T = S$, while the conservation of flux in a lossless medium implies $S^\dagger S = 1$. The polar decomposition of the scattering matrix [ ,] gives

$$S = \begin{bmatrix} V_s & 0 \\ 0 & U_s \end{bmatrix} \begin{bmatrix} -\sqrt{1-\tau} & \sqrt{\tau} \\ \sqrt{\tau} & \sqrt{1-\tau} \end{bmatrix} \begin{bmatrix} V_s^T & 0 \\ 0 & U_s^T \end{bmatrix} = \begin{bmatrix} -V_s\sqrt{1-\tau}V_s^T & V_s\sqrt{\tau}U_s^T \\ U_s\sqrt{\tau}V_s^T & U_s\sqrt{1-\tau}U_s^T \end{bmatrix},$$
(11)

where $V_s$ and $U_s$ are unitary matrices. Here,

$$t = U_t\sqrt{\tau}V_t^\dagger = U_s\sqrt{\tau}V_s^T = U_s\sqrt{\tau}(V_s^*)^\dagger, \text{ so } U_t = U_s, V_t = V_s^*$$
(12a)

$$t' = U_{t'}\sqrt{\tau}V_{t'}^{\dagger} = V_s\sqrt{\tau}U_s^T = V_s\sqrt{\tau}(U_s^*)^{\dagger}, \text{ so } U_{t'} = V_s, V_{t'} = U_s^*$$
(12b)
$$r = U_r\sqrt{\tau}V_r^{\dagger} = -V_s\sqrt{1-\tau}V_s^T = -V_s\sqrt{1-\tau}(V_s^*)^{\dagger}, \text{ so } U_r = -V_s, V_r = V_s^*$$
(12c)
$$r' = U_{r'}\sqrt{1-\tau}V_{r'}^{\dagger} = U_s\sqrt{1-\tau}U_s^T = U_s\sqrt{1-\tau}(U_s^*)^{\dagger}, \text{ so } U_{r'} = U_s, V_{r'} = U_s^*$$
(12d)

With these relations, and the knowledge of $\tau$, $U_t$ and $V_t$, we can obtain $t', t, r'$ and, as thus the full scattering matrix. In the main text, which deals primarily with the TM, we have indicated $U_t$ by $U$ and $V_t$ by $V$. Thus, for the transmission matrix, the incident singular vector is $V_t$, and the transmitted singular vector is $U_t$. For the reflection matrix, the incident eigenvector is $V_t$, the reflected singular vector is $-V_t^*$.

The velocity of the $n^{\text{th}}$ transmitted eigenchannel is $\text{v}_{n,t} = \sum_{m=1}^{N} \text{v}_m |V_{t,n,m}|^2$, the velocity of the $n^{\text{th}}$ reflected eigenchannel is $\text{v}_{n,r} = \sum_{m=1}^{N} \text{v}_m |-V_{t,n,m}^*|^2 = \text{v}_{n,t}$.

## 5. Energy density at the input surface

Here we show that the average energy density at the input surface is equal to the sum of averages of the energy densities of the incident and reflected waves, $u_n(0) = u_{n,i}(0) + u_{n,r}(0)$. This is a consequence of the proportionality of the reflected transmission eigenchannel and the complex conjugate of the incident TE, which was shown in Appendix D. Since the demonstration is based on relationships in single configurations, we distinguish between a property of a single configuration and the average over configurations by using a bracket for the latter.

Since the reflected transmission eigenchannel is proportional to the complex conjugate of the incident TE, the longitudinal velocities of the incident and reflected transmission eigenchannels in each sample configuration are the same, $\text{v}_{n,i} = \text{v}_{n,r}$. The linear energy density in a transmission eigenchannel at the sample input for a single sample configuration normalized to the spatial average of the energy density over all transmission eigenchannels can therefore be expressed as an amplitude squared of a sum of waves traveling to the right, $A_n(0)$, and left,

$$B_n(0), u_n(0)\text{v}_+ = |A_n(0) + B_n(0)|^2, \quad (13)$$

with $B_n(0) = r_n A_n^*(0)$. Here $r_n$ is the field reflection coefficient for the $n^{\text{th}}$ transmission eigenchannel in a single configuration, with $r_n^2 = 1 - \tau_n$. We can, thus, write

$$A_n(0) + B_n(0) = A_n(0) + r_n A_n^*(0). \quad (14)$$

Writing $A_n(0) = r_n A_n(0) + (1 - r_n)A_n(0)$, leads to
$A_n(0) + B_n(0) = r_n(A_n(0) + A_n^*(0)) + (1 - r_n)A_n(0) = 2r_n \text{Re}(A_n(0)) + (1 - r_n)A_n(0)$.
Averaging $u_n(0)\text{v}_+ = (A_n(0) + B_n(0))(A_n^*(0) + B_n^*(0))$ over a random ensemble gives,
$\langle u_n(0)\rangle \text{v}_+ = \langle 4r_n^2[\text{Re}(A_n(0))]^2 - (1-r_n)^2|A_n(0)|^2 + 4r_n[\text{Re}(A_n(0))]^2(1-r_n)\rangle$.

Since, $\langle Re(A_n(0))^2 \rangle = \langle Im(A_n(0))^2 \rangle$, $\langle |A_n(0)|^2 \rangle = 2\langle Re(A_n(0))^2 \rangle$. This gives,
$\langle u_n(0) \rangle v_+ = \langle [Re(A_n(0))]^2 4r_n^2 - [Re(A_n(0))]^2 2(1-r_n)^2 + [Re(A_n(0))]^2 4r_n(1-r_n) \rangle$.
Thus,
$\langle u_n(0) \rangle v_+ = \langle [Re(A_n(0))]^2 [4r_n^2 - 2(1-r_n)^2 + 4r_n(1-r_n)] \rangle = \langle [Re(A_n(0))]^2 [2(1+r_n^2)] \rangle$.
But, $\langle 2(1+r_n^2) \rangle = 2\langle 1 + (1-\tau_n) \rangle = 2\langle 2-\tau_n \rangle$. Hence,
$\langle u_n(0) \rangle v_+ = \langle 2[Re((1-\tau_n))]^2 (2-\tau_n) \rangle = \langle |A_n(0)|^2 (1+(1-\tau_n)) \rangle$.
Finally, $\langle u_n(0) \rangle v_+ = \langle u_{n,i}(0) \rangle v_+ + \langle u_{n,r}(0) \rangle v_+$, so that
$$\langle u_n(0) \rangle = \langle u_{n,i}(0) \rangle + \langle u_{n,r}(0) \rangle. \qquad (15)$$

## 6. Normalization of energy density profile by its spatial average
Energy density profiles for ensembles with different values of $g$ are shown in Fig. 6(a). The energy density profile $u(z)$ is normalized by its average over configuration and position. This average is obtained by noting that the average energy densities excited in each transmission eigenchannel on the opposite side of the sample for excitation from the left, $u_n(L)$, and the right, $u'_n(0)$, are equal, $u'_n(0) = u_n(L) = \frac{\tau_n}{v_n}$. Thus, the sum of energy density at the input for excitation in all channels on both sides of the sample is $U(0) = \sum_{n=1}^{N}[u_n(0) + u'_n(0)] = \sum_{n=1}^{N} u(0) + u(L)$. Using Eq. (2), gives, $U(0) = \frac{2N}{v_+}$. Together with the result above that $\rho_{\omega,L} = \frac{N}{\pi v_+}$, this gives $U(0) = \pi \rho_{\omega,L}/2^{4,5}$. Since the spatial average of the energy density excited by all channels from either the left or right are equal,
$$\langle u(z) \rangle_z = \langle u'(z) \rangle_z = \frac{1}{2} U(z) = \frac{N}{v_+}. \qquad (16)$$

## 7. Expression for nonlocal diffusion coefficient
Substituting the expression for $g$ in Eq. (6), $g = \frac{2Nz_b}{L+2z_b} \frac{v_T}{v_+}$, into the expression for $D$ in Eq. (4b) gives

$$D = \frac{\frac{gL}{2}}{\frac{N}{v_+} - \frac{g}{v_T}} = \frac{\frac{2Nz_b v_T L}{L+2z_b v_+ + 2}}{\frac{N}{v_+} - \frac{2Nz_b v_T}{L+2z_b v_+}}. \qquad (17)$$

Straightforward manipulation gives,

$$D = \frac{\frac{z_b}{L+2z_b} \frac{v_T}{v_+} L}{\frac{1}{v_+} - \frac{2z_b}{L+2z_b} \frac{v_T}{v_+}} = \frac{\frac{z_b}{L+2z_b} \frac{v_T}{v_+} L}{\frac{v_T}{v_+} - \frac{2z_b}{L+2z_b} \frac{v_T}{v_+}} v_T$$

$$= \frac{\frac{z_b}{L+2z_b} L}{1 - \frac{2z_b}{L+2z_b}} v_T = \frac{z_b L}{L+2z_b - 2z_b} v_T = z_b v_T,$$

which is the result of Eq. (7).

## 8. Determining the scattering mean free path

Since the scale of disorder in the scattering model considered in simulations is much smaller than the wavelength, $a = \lambda/2\pi$, scattering is nearly isotropic, and resonances do not affect the energy velocity. We therefore expect the scattering mean free time, $\tau_s$, to be nearly independent of the direction of propagation of the wave. This is confirmed in the analysis below, which shows that the scattering mean free times for different incident waveguide modes, which have different transverse velocities, are identical.

The scattering mean free path $\ell_s$ in a sample with $\Delta\varepsilon = 0.3$, $N = 64$ and width $W = 64.5\frac{\lambda_0}{2}$ is found from the decay of the coherent flux vs. ballistic time delay of the waveguide modes. The scaling of the coherent flux of waveguide modes, $|\langle t_{mm}(L)\rangle|^2$, is shown in Supplementary Fig. 4(a). $|\langle t_{mm}(L)\rangle|^2$ is the ensemble average of the flux of the $m^{th}$ waveguide mode at the output of a random sample of length $L$. The waveguide modes with higher index, $m$, which have smaller group velocities, $v_{wm}$, decay more rapidly. The decay of $|\langle t_{mm}\rangle|^2$ vs. coherent time delay $L/v_{wm}$, is shown in Supplementary Fig. 4(b). The decay times of the coherent flux for the $m^{th}$ waveguide mode is[6], $\tau_{s,m} = \frac{\ell_{s,m}}{v_m}$. Since all the plots collapse to a single curve and fall exponentially to give a single mean free scattering time $\tau_s$, and the speed of the wave in the medium is c, the scattering mean free path is $\ell_s = c\tau_s = 16.0$ m.

## 9. Pulse propagation

Under steady-state illumination, $D(W, L)$ can be expressed via Fick's first law in terms of the parameters of the TEs, $\tau_n$ and $v_n$, as in Eq. (4a). In the time domain, $D(W, L)$ is given by Fick's second law as the ratio of the partial derivative of concentration with time and the Laplacian of the concentration. In samples thicker than five times the mean free path, the transmitted intensity in a diffusive medium falls exponentially in time once higher order diffusion modes have decayed, with a decay rate $1/\tau_D = \pi^2 D/(L + 2z_b)^2$ [7–9,8]. This is the decay rate of the quasi-normal modes or resonances of the medium and equals the linewidth of quasi-normal modes [10,11].

The transmission due to an incident Gaussian pulse is obtained by multiplying field transmission spectra with the Gaussian spectrum of the envelope for the field pulse and Fourier transforming into the time domain. The ensemble average of the transmitted intensity for the sample for which steady-state results are shown in Fig. 4(a) with $N = 64$ and $L = 7\ell_s$, at which $\frac{D}{\widetilde{D_0}}$ reaches its peak value of 0.91, is shown in Fig. 4(a). The transmitted pulse decays exponential with decay time $\tau_D = 0.68$ μs $\pm 0.01$. Beyond a delay of 5 μs, the decay rate slows due to the increasing relative weight of longer lived quasi-normal modes[10–12]. The diffusion coefficient given by $D = \frac{(L+2z_b)^2}{\pi^2 \tau_D}$[9], which is equal to $2.16 \times 10^9 \pm 0.02$ m²/s, obtained with $z_b$ taken as its average value over the spectrum for the incident Gaussian pulse, $z_b = 11.4$. The average scattering mean free path over the spectrum of the incident pulse of $\ell_s = 16.0 \pm 0.15$ m, gives $\widetilde{D_0} = \frac{1}{2}c\ell_s = 2.40 \times 10^9$ m²/s and $D/\widetilde{D_0} = 0.90 \pm 0.015$, in agreement with the state-diffusion coefficient for this sample, as seen in Fig. 4(a). The uncertainty in $D/\widetilde{D_0}$ is due to different values of the slope of the curve in the inset obtained for different time ranges.

Though the diffusion coefficients found in steady-state and in time-domain simulations agree, still $D$ falls 10% below $\widetilde{D_0}$. Several factors contribute to this difference. At the length at which $D(L)$ reaches its peak value, the diffusion coefficient is already lowered below the bare diffusion coefficient due to weak localization[10,11,13]. The bare diffusion coefficient might be obtained by extrapolating the linear region of $D(L)/\widetilde{D_0}$ in Fig. 4(a) back to $L = 0$. This gives, $D(0)/\widetilde{D_0} = 0.95$. In the diffusive regime, the length dependent diffusion coefficient, $D(L)$, falls linearly with length from the bare diffusion coefficient with the fractional reduction in $D(L)$ proportional to $\frac{1}{g_{Th}} \sim \tau_{Th}/\tau_T$ because of the increasing probability of a path looping back upon a coherence length of the path with increasing sample thickness [10]. Similarly, the time dependent diffusion coefficient proportional to the decay rate of $\ln\langle I(t)\rangle$ falls linearly with delay time, with the drop in decay rate proportional to $t/\tau_T$[10,11].

Another reason that $D/\widetilde{D_0}$ may fall below unity is that $\widetilde{D_0} = \frac{1}{2}c\ell_s$ may be larger than the bare diffusion coefficient $D_0 = \frac{1}{2}v_E\ell$[14]. There are two countervailing factors, $\ell$ may be larger than $\ell_s$, while $v_E$ may be smaller than $c$. These corrections are likely to be small since $a = \lambda_0/2\pi$, so that the individual scattering elements are too small to support resonances so that $v_E \sim c$, and scattering might be nearly isotropic with $\ell \sim \ell_s$. However, it should be noted that the prevailing assumption that scattering is isotropic and is independent of sample dimensions, which leads to the classical result for the diffusion coefficient, is shown here not to hold for wave diffusion in bounded media.

## 10. Scaling of diffusion coefficient and its factors

The scale-dependent diffusion coefficient $D(W,L)$ relative to $\widetilde{D_0} = \frac{1}{2}c\ell_s$ and the factors of the diffusion coefficient $z_b(W,L)$ and $v_T(W,L)$, as given in Eq. (7), are given in Supplementary Fig. 6. Results for samples with disorder $\Delta\varepsilon = 0.3$ and for $N = 8$ and $N = 64$ are shown, respectively, in the top and bottom rows over the full range of $W$ for each $N$ and for a wide range of $L/\ell_s$. The variation with width is smaller for larger $N$, but is still substantial for $N = 64$.

## 11. Breakdown of diffusion in energy density profile

We find that energy density within the sample falls linearly with depth for sample with lengths nearly equal to the localization length $L = \xi$ and $g = 1$. Here we consider the variation of energy density in an ensemble of samples with $g = 1.078$. The sample width is $W = (N + \frac{1}{2})\frac{\lambda_0}{2} = 8.5\frac{\lambda_0}{2}$. The derivative of the normalized energy density inside the medium is shown as the blue curve in Supplementary Fig. 7(a). Since the position dependent diffusion coefficient and the derivative of the energy density depend upon the distance from the nearest boundary, $(du(z)/dz)\frac{v_+}{N}$ is symmetric about the center of the sample. To second order in the displacement from the center of the sample, $(du(z)/dz)\frac{v_+}{N} = a + b\left[(z - \frac{L}{2})/(L/2)\right]^2$, where $a = \frac{d\langle u(z)\rangle}{dz}|_{z=L/2}\frac{v_+}{N}$ and $b = \frac{d^2\langle u(z)\rangle}{dz^2}|_{z=\frac{L}{2}}\frac{v_+}{N}$. The fit of this function to $(du(z)/dz)\frac{v_+}{N}$ gives $a = -1.826$, and $b = 0.177$, and is displayed as the red curve. $u(z)\frac{v_+}{N}$ found in simulations (blue curve) and computed with use of the parameters $a$ and $b$ found in (a) (red curve) are seen to overlap in Supplementary Fig. 7(b).

**Supplementary References**